\documentclass[%
reprint,
 amsmath,amssymb,
pra,
]{revtex4-2}
\usepackage{amsmath}
\usepackage{graphicx}
\usepackage{dcolumn}
\usepackage{bm}
\usepackage{hyperref}
\hypersetup{colorlinks=true, citecolor=blue, urlcolor=blue, linkcolor=blue}
\usepackage{subfigure}

\begin{document}


\title{Gap solitons of the Wannier and Bloch types in spin-orbit-coupled Bose-Einstein condensates with a moir\'{e} lattice}

\author{Jun-Tao~He\textsuperscript{1}}
\author{Xue-Ping~Cheng\textsuperscript{2}}
\author{Xin-Wei~Jin\textsuperscript{1}}
\author{Hui-Jun~Li\textsuperscript{1}}
\author{Ji~Lin\textsuperscript{1}}%
\thanks{Corresponding author. Email: linji@zjnu.edu.cn}
\author{Boris A. Malomed\textsuperscript{3,4}}
\thanks{Corresponding author. Email: malomed@tauex.tau.ac.il}
\affiliation{\textsuperscript{1}Department of Physics, Zhejiang Normal University, Jinhua 321004, China}
\affiliation{\textsuperscript{2}School of Science, Zhejiang University of Science and Technology, Hangzhou 310023, China}
\affiliation{\textsuperscript{3}Department of Physical Electronics, Faculty of Engineering, and Center for Light-Matter Interaction, Tel Aviv University, Tel Aviv 69978, Israel}
\affiliation{\textsuperscript{4}Instituto de Alta Investigaci\'{o}n, Universidad de Tarapac\'{a}, Casilla 7D, Arica 1000000, Chile}
%


\begin{abstract}
	Gap solitons (GSs) bifurcating from flat bands, which may be represented in terms of Wannier functions, have garnered significant interest due to their strong localization with extremely small norms. Moir\'{e} lattices (MLs), with multiple flat bands, offer an appropriate platform for creating such solitons. We explore the formation mechanism and stability of GSs in spin-1 Bose-Einstein condensates under the combined action of the Rashba spin-orbit coupling (SOC) and an ML potential. We identify five Wannier-type GS families bifurcating from the lowest five energy bands in the spectrum induced by the ML with sufficiently large period and depth. These fundamental GSs serve as basic elements for constructing more complex Wannier-type GS states. Reducing the lattice period and depth triggers a transition from the Wannier-type GSs to those of the Bloch type, the latter exhibiting higher norm thresholds and pronounced spatial broadening near edges of the energy bands. In addition to tuning the lattice-potential parameters, adjusting the SOC strength can also modulate the flatness of energy bands and enhance the localization of gap solitons, enabling reversible transitions between the GSs of the Wannier and Bloch types. Distinctive properties of GSs in the quasiperiodic ML are uncovered too. Thus, we propose the theoretical foundation for the creation of and manipulations with strongly localized GSs.
\end{abstract}

\maketitle

\section{Introduction}
Gap solitons (GSs) are nonlinear localized states populating spectral gaps induced by spatially periodic or quasiperiodic lattice potentials. GSs have been extensively studied in many fields, especially nonlinear optics and Bose-Einstein condensates (BECs) \cite{KartashovYV2011Solitons,PelinovskyDE2011Localization,DongL2010Shaping}. In particular, in two dimensions (2D) families of GS modes have been studied under the action of various potentials representing triangular \cite{KevrekidisPG2002Solitons,AlexanderTJ2007Multivortex}, square-shaped \cite{OstrovskayaEA2003Matter,ZhangY2015Gap,LiJ2021Dark}, hexagonal \cite{PelegO2007Conical,MengH2023Vector}, ring-shaped \cite{BaizakovBB2006Matter,KartashovYV2019Stable}, and quasiperiodic \cite{SakaguchiH2006Gap}. GSs are classified into two distinct types according to their shapes. A prevalent one is the Bloch type (BT) \cite{ShiZ2007Solitary}. It features widely expanding GS wave functions, which asymptotically approach the corresponding linear Bloch states when the GS is created near edges of the corresponding energy bands that are adjacent to the spectral gap hosting the GS. The BT GSs exist above a large norm threshold and are challenging to excite. GSs of the other type bifurcate from nearly flat bands and can be excited even with a very small norm. Being approximated by Wannier functions rather than the Bloch ones, they are referred to as Wannier-type (WT) GSs \cite{AlfimovGL2002Wannier,WangC2023Wannier}. Owing to their tight localization and dynamic stability, the WT species of GSs offers applications to quantum information and topological quantum computing \cite{WeedbrookC2012Gaussian,SuSL2020Nondestructive}.

As said above, the emergence of WT GSs is intrinsically linked to the existence of flat bands in the respective linearized system. For conventional lattices, realizing such flat bands typically requires extremely deep potential wells, yielding only a limited number of effectively flat bands and constraining further advancement in this field. Recently, moir\'{e} lattices (MLs), built as a superposition of two conventional 2D lattices with a special rotation (twist) angle between them, have drawn much interest, revealing intriguing phenomena, such as unconventional superconductivity \cite{CaoY2018Unconventional}, fractal energy spectra \cite{DeanCR2013Hofstadter}, and the localization-delocalization transition \cite{WangP2020Localization,FuQ2020Optical}. MLs exhibit a distinctive band-gap structure characterized by the alternation of multiple flat bands and wide gaps, providing a unique platform for exploring novel species of WT GSs. Initial studies of GSs in MLs were primarily conducted in terms of photonic systems, where the MLs are imprinting in photorefractive crystals by optical induction \cite{FleischerJW2003Observation}. Various types of GSs have been thus identified, including multifrequency \cite{KartashovYV2021Multifrequency}, multipole \cite{LiuX2023Gap,ZengL2024Solitons}, and vortex \cite{IvanovSK2023Vortex} solitons. With the realization of tunable twisted-bilayer optical lattices in BECs \cite{MengZ2023Atomic}, the studies of ML-supported GSs have been extended to single-component \cite{LiuX2023Matter} and two-component \cite{TuP2025Vector,WangL2025Gap} matter-wave systems. Most current studies focus on GSs bifurcating from the lowest-energy band, while systematic investigations are lacking for GS modes that may bifurcate from higher energy bands.

The spin-orbit coupling (SOC) introduces an additional tool for the work with matter-wave systems. Mathematically represented by the first-order differential operator, SOC shows the promise for controlling flat bands \cite{ZhangY2013Bose,HeJ2023Stationary,LuoHB2024Energy}. In other contexts, SOC suppresses the collapse of 2D \cite{SakaguchiH2016Vortex,SakaguchiH2018One,KartashovYV2020Multidimensional,DengH2024Semivortex} and 3D \cite{ZhangYC2015Stable} solitons, also enabling the creation of diverse types of \textquotedblleft exotic" solitons \cite{SinhaS2011Trapped,XuZF2012Symmetry,GautamS2018Three,AdhikariSK2021Multiring,GuoY2024Stable,FangP2024Soliton}. However, specific forms and properties of GSs were not addressed under the combined effect of SOC and MLs, especially in three-component BEC systems with many tunable parameters.

In this work we focus on a effectively 2D matter-wave system with SOC of the Rashba type \cite{BychkovYA1984Oscillatory,BindelJB2016Probing} applied along with an ML. For periodic MLs, we systematically investigate the existence and stability of GSs bifurcating from the lowest five energy bands for different interaction strengths and lattice periods. The interaction determines whether the GS bifurcates to the gap above or below the corresponding flat band, while the lattice period determines the type of the emerging GS. For the large lattice period, the lowest five energy bands are effectively flat, and we find five fundamental WT GS families bifurcating from these flat bands. The coherent superposition of the fundamental solitons enables the creation of more complex WT-GS modes. Conversely, for the small period, BT GSs emerge, with phase profiles congruent to those of WT GSs. We find that SOC plays a crucial role in flattening the energy bands, thereby facilitating the transition of WT GSs to BT GSs. Finally, we extend the analysis to quasiperiodic MLs.

The presentation is structured as follows. In Sec.~\ref{4-2} we describe the mean-field model for the spin-1 BEC with the Rashba SOC loaded into the ML. In Sec.~\ref{4-3} we report systematically produced analytical and numerical results for periodic and quasiperiodic MLs, including the linear bandgap spectrum, two (WT and BT) GSs species, and effects of various parameters. The paper is concluded in Sec.~\ref{4-4}.

\section{Model}\label{4-2}
We consider a quasi-2D spin-1 BEC, with coordinates $\left(x,y\right)$, under the action of the Rashba SOC and ML. In the framework of the mean-field approximation, the system is governed by the three-component Gross-Pitaevskii equation \cite{KawaguchiY2012Spinor,GautamS2017Vortex}, which is written here in the scaled form:
\begin{equation}\label{4eq1}
\begin{aligned}
i\partial_t\psi_{\pm 1}=&\left(-\frac{1}{2} \nabla^2+V\right)\psi_{\pm 1}-\frac{\gamma}{\sqrt{2}}\partial_{\mp}\psi_0+c_2 \psi_0^2 \psi_{\mp 1}^*\\
&+\left[c_0 \rho+c_2\left(\rho_{\pm 1}+\rho_0-\rho_{\mp 1}\right)\right] \psi_{\pm 1},\\
i\partial_t \psi_0=&\left(-\frac{1}{2} \nabla^2+V\right)\psi_0+\left[c_0\rho+c_2\left(\rho_{+1}+\rho_{-1}\right)\right] \psi_0\\
&+2 c_2 \psi_0^* \psi_{+1} \psi_{-1} -\frac{\gamma}{\sqrt{2}}\left(\partial_{+}\psi_{+1}+\partial_{-}\psi_{-1}\right),
\end{aligned}
\end{equation}
where the units of the energy, time, and length are $\hbar \omega_z$, $\omega_z^{-1}$, and $l_z=\sqrt{\hbar/(m\omega_z)}$, respectively. $m$ is the atomic mass, and $\omega_z$ is the trapping frequency applied to the BEC layer in the transverse direction. $\psi_j$ and $\rho_j=\left|\psi_j\right|^2 $ (with $j=\pm 1,0$) are three-component wave functions and densities, respectively. $\rho=\sum_j \left|\psi_j\right|^2$ is the total density with the norm $N$. Constants of the mean-field and spin-exchange interactions are $c_0=2\sqrt{2\pi}\mathcal{N}(a_0+2a_2)/(3Nl_z)$ and $c_2=2\sqrt{2\pi}\mathcal{N}(a_2-a_0)/(3Nl_z)$, respectively, where $a_0$ and $a_2$ are two-body \textit{s}-wave scattering lengths for the total spin $0$ and $2$ \cite{GautamS2015Analytic}. $\mathcal{N}$ is the number of atoms.  Further, $\gamma$ is the strength of Rashba SOC, which is represented by operators $\partial_{\pm}=i\partial_y\pm \partial_x$. The ML is represented by its potential, constructed by overlapping two square-shaped optical lattices with a twist angle $\theta$ between them:
\begin{equation}\label{4eq2}
\begin{aligned}
V(x,y)=&V_0\left[\sin^2(\frac{\pi}{a}x_+)+\sin^2(\frac{\pi}{a}y_+)\right]
\\&+V_0\left[\sin^2(\frac{\pi}{a}x_-)+\sin^2(\frac{\pi}{a}y_-)\right],
\end{aligned}
\end{equation}
where $a$ and $V_0$ are the period and depth of the sublattices, and $(x_{\pm},y_{\pm})$ are produced by the rotating the original coordinates:
\begin{equation}
\left(\begin{aligned}x_{\pm}\\y_{\pm}\end{aligned}\right)=
\left(\begin{aligned}&\cos(\frac{\pi}{4}\pm\frac{\theta}{2}), &-\sin(\frac{\pi}{4}\pm\frac{\theta}{2})\\
&\sin(\frac{\pi}{4}\pm\frac{\theta}{2}), &\cos(\frac{\pi}{4}\pm\frac{\theta}{2})\end{aligned}\right)
\left(\begin{aligned}x\\y\end{aligned}\right).
\end{equation}
When angle $\theta$ is a Pythagorean angle of a right triangle with integer sides ($\bar{a}^2+\bar{b}^2=\bar{c}^2$, where $\bar{a}$, $\bar{b}$, and $\bar{c}$ are positive integers), the two sublattices form a periodic ML $V(x,y)$ with the period $\sqrt{\bar{c}}a$. Otherwise, $V(x,y)$ is a quasiperiodic ML.

In the experiment the Rashba SOC can be induced by the Raman coupling in the square optical lattice \cite{WuZ2016Realization}. The ML system is produced by adding another square optical lattice (which is rotated by angle $\theta$). Parameters ($\gamma$, $a$, $V_0$) may be controlled by adjusting laser beams illuminating the condensate. Furthermore, the nonlinear parameters $c_0$ and $c_2$ can be adjusted by means of the Feshbach-resonance technique \cite{PapoularDJ2010Microwave,ChinC2010Feshbach}.

\section{Results and discussion}\label{4-3}
\subsection{The band-gap structures}
\begin{figure}[t]
	\centering
	\includegraphics[width=1\linewidth]{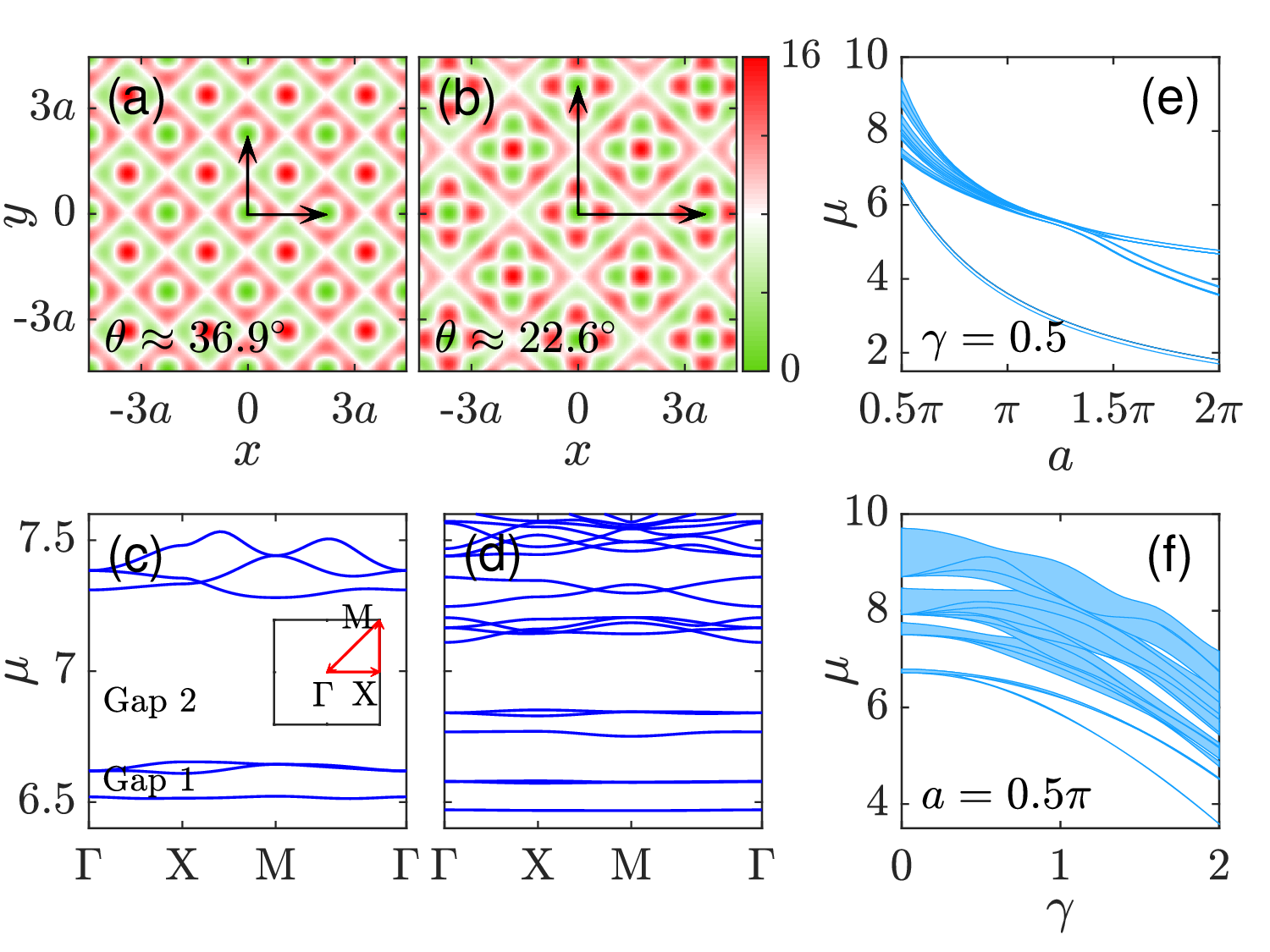}
	\caption{The ML profiles for twisted angles (a) $\theta=\arctan(3/4)\approx 36.9^{\circ}$ and (b) $\theta=\arctan(5/12)\approx 22.6^{\circ}$. The black arrows represent the lattice vectors. The band-gap structures along the high-symmetry lines for the MLs with (c) $\theta=\arctan(3/4)$ and (d) $\theta=\arctan(5/12)$ when $a=0.5\pi$ and $\gamma=0.5$. The box and red arrows in (c) represent the first Brillouin zone and the high-symmetry lines. (e,f) The effect of $\gamma$ and $a$ on the ML band-gap structure with $\theta=\arctan(3/4)$. These blue (white) areas represent energy bands (gaps). Here, $V_0=4$.}
	\label{4fig:1}
\end{figure}

When the twisted angle $\theta$ takes a Pythagorean value, the two square sublattices form moir\'e patterns with the periodic translational symmetry, the period being $\sqrt{\bar{c}}a$. We focus on two Pythagorean angles, $\theta=\arctan(3/4)$ and $\theta=\arctan(5/12)$, with the profiles of the corresponding MLs with $V_0=4$ shown in Figs.~\ref{4fig:1}(a) and \ref{4fig:1}(b), which are affected by the lattice period $a$.
The perpendicular lattice vectors (black arrows) imply that the reciprocal lattice vectors are also perpendicular to each other. The length of the reciprocal lattice vector is $2\pi/(\sqrt{\bar{c}}a)$, which is also the side width of the first Brillouin zone, as shown in the subplot of Fig.~\ref{4fig:1}(c). The ML periodicity makes it possible to calculate the linear band-gap spectrum of the system by means of the linear Bloch theory, which is the necessary step in the study of GSs.

According to the linear Bloch theory, the eigenfunctions $\phi_j$ of the linearized form of Eq.~\eqref{4eq1} are looked for as
\begin{equation}\label{4eq4}
\psi_j=e^{i(k_x x+k_y y-\mu_n t)}\phi^{(n)}_j(x,y,k_x,k_y),
\end{equation}
where $\mu_n$ is the chemical potential of the energy band with index $n$. $k_x$ and $k_y$ are the Bloch quasi wave numbers. By substituting Eq.~\eqref{4eq4} in the linearized equation \eqref{4eq1}, we derive the equations that the eigenfunctions $\phi^{(n)}_j$ satisfy:
\begin{equation}\label{4eq5}
\begin{aligned}
\mu_n\phi^{(n)}_{\pm 1} =&-\frac{1}{2}\left[\left(\partial_x+ik_x\right)^2+\left(\partial_y+ik_y\right)^2\right]\phi^{(n)}_{\pm 1}\\
&+V\phi^{(n)}_{\pm 1}-\frac{\gamma}{\sqrt{2}}\left(i\partial_y-k_y\mp\partial_x\mp ik_x\right)\phi^{(n)}_0,
\\
\mu_n\phi^{(n)}_0 =&-\frac{1}{2}\left[\left(\partial_x+ik_x\right)^2+\left(\partial_y+ik_y\right)^2\right]\phi^{(n)}_0\\
&+V\phi^{(n)}_0-\frac{\gamma}{\sqrt{2}}\left(i\partial_y-k_y+\partial_x+ik_x\right)\phi^{(n)}_{+1}\\
&-\frac{\gamma}{\sqrt{2}}\left(i\partial_y-k_y-\partial_x-ik_x\right)\phi^{(n)}_{-1}.
\end{aligned}
\end{equation}
Solving Eq.~\eqref{4eq5} by means of the Fourier collocation method, we can obtain the chemical potential as the function of the quasi wave numbers, $\mu_n(k_x,k_y)$, and the corresponding eigenfunctions $\phi_j^{(n)}(x,y,k_x,k_y)$ in the first Brillouin zone. Figures \ref{4fig:1}(c) and \ref{4fig:1}(d) show the band-gap structures along the high-symmetry lines for the MLs with $\theta=\arctan(3/4)$ and $\theta=\arctan(5/12)$, respectively. Comparing the lowest three energy bands in these two cases, we find that the energy bands for $\theta=\arctan(5/12)$ are relatively flat and have lower values of $\mu_n$. However, the energy gaps for $\theta=\arctan(3/4)$ are much wider than those for $\theta=\arctan(5/12)$, especially for the second finite energy gap. The presence of multiple and wide energy gaps enables the system to support various GS types.

The depth and period of the lattice potential significantly affect the band-gap structure, a deeper lattice potential resulting in narrower bands and wider band gaps. We consider the effect of lattice period $a$ on the band-gap structure for $\theta=\arctan(5/12)$, $V_0=4$, and $\gamma=0.5$, as shown in Fig.~\ref{4fig:1}(e). As the lattice period $a$ increases, the chemical potential $\mu$ of each energy band decreases. Simultaneously, the increase in $a$ leads, first, to a merger of the high-energy bands, followed by splitting into narrower (nearly flat) bands, accompanied by the disappearance and reemergence of the gaps which separate the bands. The emergence of multiple flat bands here is an ML-produced effect.

We also address the effect of the SOC strength $\gamma$ on the band-gap structure fixing given $\theta=\arctan(5/12)$, $V_0=4$, and $a=0.5$, as shown in Fig.~\ref{4fig:1}(f). Similar to the effect of $a$, the increase in $\gamma$ leads to a decrease in the chemical potential of each energy band, but at a faster rate. Due to the varying rates of the decrease for each energy band, the widths of the energy gaps are affected and an additional energy gap forms.

\subsection{Wannier-type and Bloch-type gap solitons}
We have found bright solitons, i.e., GSs populating the gaps. If both interaction coefficients $c_{0}$ and $c_{2}$ are negative (which corresponds to the attractive nonlinearity), the 2D system suffers the collapses, as might be expected \cite{BergeL1998Wave,FibichG2015The}, which inhibits the formation of GSs. On the other hand, if both $c_0$ and $c_2$ are positive (which corresponds to the repulsive nonlinearity), the broadening of wave functions induced by dispersion cannot be balanced by the repulsive interactions. GSs can only exist in a large-scale potential well (binding all atoms) of the ML, which is similar to the harmonic-oscillator trapping potential \cite{WangQ2024Vector}. Here, we produce GS solutions of Eq.~\eqref{4eq1} in the most interesting case of $c_0c_2<0$, i.e., two cases $c_0<0$, $c_2>0$ (antiferromagnetic), and $c_0>0$, $c_2<0$ (ferromagnetic) \cite{KawaguchiY2012Spinor}. Here, we fix $\theta=\arctan(3/4)$ to deal with broader energy gaps.

GS solutions $\varphi_j(x,y)$ satisfy the stationary-state equations, which are obtained by substituting $\psi_j(x,y,t)=\varphi_j(x,y)e^{i\mu t}$ into Eq.~\eqref{4eq1}:
\begin{equation}\label{4eq6}
\begin{aligned}
\mu \varphi_{\pm1}=&\left(-\frac{1}{2} \nabla^2+V\right)\varphi_{\pm 1}-\frac{\gamma}{\sqrt{2}}\partial_{\mp}\varphi_0+c_2 \varphi_0^2 \varphi_{\mp 1}^*\\
&+\left[c_0 \rho+c_2\left(\rho_{\pm 1}+\rho_0-\rho_{\mp 1}\right)\right] \varphi_{\pm 1},
\\
\mu \varphi_0=&\left(-\frac{1}{2} \nabla^2+V\right)\varphi_0+\left[c_0\rho+c_2\left(\rho_{+1}+\rho_{-1}\right)\right] \varphi_0\\
&+2 c_2 \varphi_0^* \varphi_{+1} \varphi_{-1} -\frac{\gamma}{\sqrt{2}}\left(\partial_{+}\varphi_{+1}+\partial_{-}\varphi_{-1}\right) .
\end{aligned}
\end{equation}
In the presence of the ML and SOC, GSs $\varphi_j(x,y)$ remain solutions of Eq.~\eqref{4eq6} under the action of some symmetric operations. In particular, the system obeys the U(1) symmetry, defined by $\hat U(\theta_0)\varphi_j(x,y)=e^{i\theta_0}\varphi_j(x,y)$, and the space-translation symmetry $\hat T(\Delta x,\Delta y)\varphi_j(x,y)=\varphi_j(x+\Delta x,y+\Delta y)$, associated with the period $\sqrt{\bar{c}}a$, where both $\Delta x$ and $\Delta y$ are integer multiples of $\sqrt{\bar{c}}a$. The system also obeys the spin-flip symmetry,
\begin{equation}\label{4eq7}
\hat O \varphi_j(x,y)=\pm(-1)^j\varphi_{-j}^*(x,y),
\end{equation}
and the axial symmetry,
\begin{equation}\label{4eq8}
\hat D\varphi_j(x,y)=\varphi_{-j}(y,x)e^{i\pi j/2}.
\end{equation}
These symmetries facilitate the analysis.

\begin{figure}[t]
	\centering
	\includegraphics[width=1\linewidth]{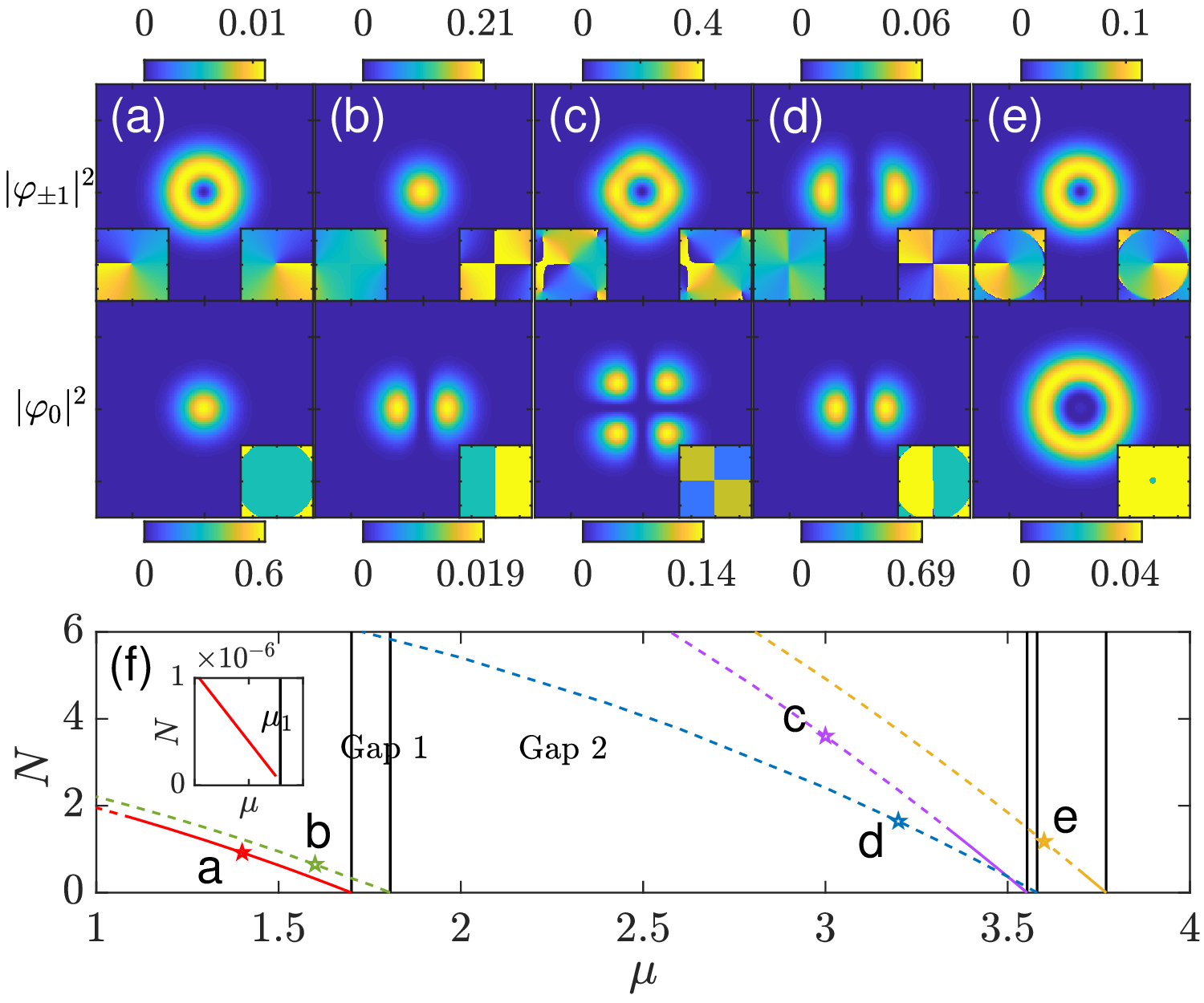}
	\caption{WT (Wannier-type) GS families bifurcating from the lowest five flat bands for the ML with $\theta=\arctan(3/4)$, $a=2\pi$, and $V_0=4$. Other parameters are $\gamma=0.5$, $c_0=-1$, and $c_2=1.5$ (the antiferromagnetic system). (a)–(e) The density profiles of three components $\varphi_{\pm1}$ (the first row) and $\varphi_0$ (the second row) for five GS families. The corresponding phases are shown in the corresponding subfigures, where the phase of $\varphi_{+1}$ is in the left subfigure. The drawing grid is $(x,y)\in [-3,3]$. (f) Norm $N$ vs the chemical potential $\mu$ for the five GS families. The solid (dashed) parts of these curves indicate that the corresponding GSs are stable (unstable). The black lines represent five flat bands with $\mu_1=1.6998$, $\mu_2=1.8065$, $\mu_3=3.5533$, $\mu_4=3.5808$, and $\mu_5=3.7703$. Here, the inset shows the norm curve of the GS family bifurcating from $\mu_1$, plotted near the band edge.}
	\label{4fig:2}
\end{figure}

GS solutions to Eq.~\eqref{4eq6} were obtained using the numerical squared-operator iteration method \cite{YangJ2010Nonlinear}, with the computational grid $(x,y)\in[-12\pi,+12\pi]$ and the number of discrete points is $1024\times 1024$. For the ML with $\theta=\arctan(3/4)$, $a=2\pi$, and $V_0=4$, the low-energy bands exhibit flat dispersion, and several gaps are present. We have found five types of GSs, setting $\gamma=0.5$, $c_0=-1$, and $c_2=1.5$ (the antiferromagnetic system). The density profiles of the five GS types are plotted in Figs.~\ref{4fig:2}(a)-\ref{4fig:2}(e), featuring ring-shaped and multi hump patterns.
Note that the density profiles of the GS components $\psi_{+1}$ and $\psi_{-1}$ are identical, but they have different phases. In Fig.~\ref{4fig:2}(f) we also present the relation between the soliton's norm, $N=\int_{-\infty}^{+\infty}\sum_j |\varphi_j|^2\mathrm{d}x\mathrm{d}y$, and the chemical potential $\mu$ for the five GS families. The norm curves bifurcate at the chemical potential $\mu_n$ ($n=1,2,3,4,5$) of the corresponding flat bands and pass through the lower-energy flat bands toward $\mu \rightarrow -\infty$. In other words, the chemical potential of the GS families is
\begin{equation}\label{4eq9}
\mu=\mu_n+\Delta\mu,
\end{equation}
where $\Delta\mu<0$, provided that $c_0<0$. It is easy to understood that the offset $\Delta\mu$ is determined by $c_0$ because these GS solutions are invariant with respect to the spin-flip symmetry \eqref{4eq7}, which implies that the nonlinear terms associated with $c_2$ are eliminated when substituting the GS solutions into Eq.~\eqref{4eq6}. As $\mu$ increases, the density profiles of these five GS types remain almost unchanged (slightly widened), but norm $N\rightarrow 0$ as $\mu\rightarrow\mu_n$. Here, the norm of GSs exhibits a very small threshold on the order of $10^{-8}$, as shown in the inset of Fig. \ref{4fig:2}. In addition, the square minimal unit cell of the ML gives rise to three other flat bands that are very close to those ($\mu_2,\mu_3,\mu_4$), and there exist three additional GS types (not shown here) whose density profiles are very similar to those plotted in Figs.~\ref{4fig:2}(b)-\ref{4fig:2}(d), rotated by $\pi/4$. The norm curves of the two degenerate GS types are strongly overlapping. In fact, all these GSs are similar to Wannier functions $w_j^{(n)}(x,y)$ (in the central unit cell) corresponding to $\mu_n$, especially in the limit of small amplitudes, where the nonlinear terms may be treated as small perturbations, i.e.,
\begin{equation}\label{4eq10}
\varphi^{(n)}_j\approx w_j^{(n)}(x,y)=\frac{1}{S}\int_{\mathrm{BZ}}\phi_j^{(n)}(x,y,k_x,k_y)\mathrm{d}k_x\mathrm{d}k_y,
\end{equation}
where $S=\pi^2/(5a^2)$ is the area of the first Brillouin zone \cite{ShenS2024Two}. The Wannier functions corresponding to the flat bands are eigenfunctions of the linear part of Eq.~\eqref{4eq1}, with the eigenvalue of $\mu_n$. This is why these GS families bifurcate at the chemical potential $\mu_n$ corresponding to the flat bands. The GSs, which remain localized even at small values of the norm, can be well approximated by Wannier functions; therefore they are termed WT GSs.

\begin{figure}[t]
	\centering
	\includegraphics[width=1\linewidth]{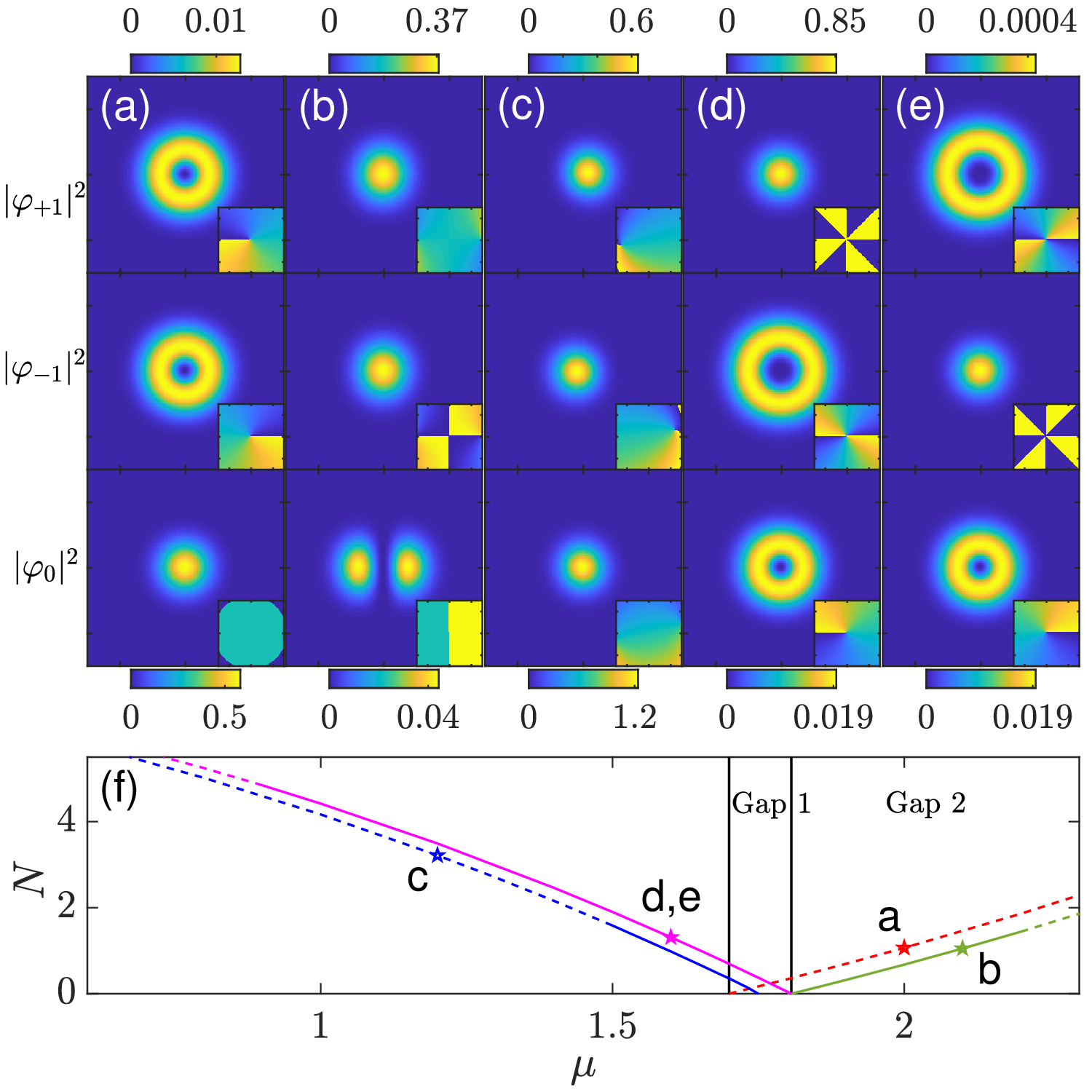}
	\caption{WT (Wannier-type) GS families bifurcating from the lowest two flat bands for the ML with $\theta=\arctan(3/4)$, $a=2\pi$, and $V_0=4$. Other parameters are $\gamma=0.5$, $c_0=1$, and $c_2=-1.5$ (the ferromagnetic system). (a)–(e) The density profiles of three components $\varphi_{+1}$ (the first row), $\varphi_{-1}$ (the second row), and $\varphi_{0}$ (the third row) for five GS families. The corresponding phases are shown in the subfigures. The first and second rows in (a)–(c) share the same color bars. $|\varphi_{-1}|$ in (d) and (e) are equal to $|\varphi_{+1}|$ in (e) and (d), respectively. The drawing grid is $(x,y)\in [-3,3]$. (f) Norm $N$ vs the chemical potential $\mu$ for the five GS families. The norm curves for (d) and (e) are identical. The solid (dashed) parts of these curves indicate that the corresponding GSs are stable (unstable). The black lines represent two flat bands with $\mu_1=1.6998$ and $\mu_2=1.8065$.}
	\label{4fig:3}
\end{figure}

Next we address the case of $c_0=1$ and $c_2=-1.5$ (the ferromagnetic system), fixing other parameters as above, $a=2\pi$, $V_0=4$, and $\gamma=0.5$. We again find that the several types of WT GSs previously mentioned still exist and also bifurcate at the chemical potential $\mu_n$. Two examples of the GSs bifurcating from the two lowest flat bands ($\mu_1$ and $\mu_2$) are presented in Figs.~\ref{4fig:3}(a) and \ref{4fig:3}(b).
Comparing the norm curves of these two GS families [$\varphi_j^{(1)}$ and $\varphi_j^{(2)}$] for different interaction types, antiferromagnetic and ferromagnetic [the red and green lines in Figs.~\ref{4fig:2}(f) and \ref{4fig:3}(f)], we notice the difference that the GSs exist in the interval $\mu>\mu_n$ when $c_0>0$, i.e., $\Delta\mu>0$ due to the decisive role of $c_0$. Moreover, we also find three other GS types, which have different density profiles of components $\varphi_{+1}$ and $\varphi_{-1}$. In Fig.~\ref{4fig:3}(c) the GS family with different center positions of three components reveals the phase separation, cf. Ref. \cite{GuiZ2023Spin}. The two degenerate GS families in Figs.~\ref{4fig:3}(d) and \ref{4fig:3}(e) represent vortex states with a winding number difference of $1$ between the components \cite{LiuK2022Spin}. These are \textquotedblleft exotic" quantum states induced by SOC. For these GS families with $|\varphi_{+1}|^2\neq |\varphi_{-1}|^2$, the interaction terms with $c_2$ dominate over those with $c_0$; hence $c_2$ determines the offset of the chemical potential, producing $\Delta\mu<0$ ($c_2<0$), as indicated by the norm curves (the blue and magenta lines) in Fig.~\ref{4fig:3}(f). In fact, the three GS families can be approximated by a linear combination of two Wannier functions $w_j^{(1)}$ and $w_j^{(2)}$, which are, essentially, the superpositions of two corresponding GS families (with the same $\mu$) in Figs.~\ref{4fig:3}(a) and \ref{4fig:3}(b). Using numerical data, these relations can be written as
\begin{equation}\label{4eq11}
\begin{aligned}
&\varphi_j^{(\mathrm{c})}\approx A(c_0,c_2)\left[\varphi_j^{(1)}(x,y)+i\hat D\varphi_j^{(2)}(x,y)\right],\\
&\varphi_j^{(\mathrm{d,e})}\approx A(c_0,c_2)\left[\varphi_j^{(2)}(x,y)\pm i\hat D\varphi_{j}^{(2)}(x,y)\right],
\end{aligned}
\end{equation}
and the chemical potentials at the bifurcation points are $\mu_{\mathrm{c}}\approx(\mu_1+\mu_2)/2$ and $\mu_{\mathrm{d,e}}\approx\mu_2$, respectively. Here, $\hat D$ is the operator defined in Eq.~\eqref{4eq8}, and $A(c_0,c_2)$ is a coefficient determined by the nonlinearity, so that $A(c_0,c_2)\approx 1$ between the cases of $c_0=-1$, $c_2=1.5$ and $c_0=1$, $c_2=-1.5$. The three GS types also belong to WT. Similarly, the system supports various WT GSs approximately represented by other linear combinations, which are not further elaborated here. In the case of $c_0=-1$ and $c_2=1.5$, the WT GS families (approximated by the superposition of two Wannier functions) still exist and bifurcate to the right ($\Delta\mu>0$) of the bifurcation point.

\begin{figure}[t]
	\centering
	\includegraphics[width=1\linewidth]{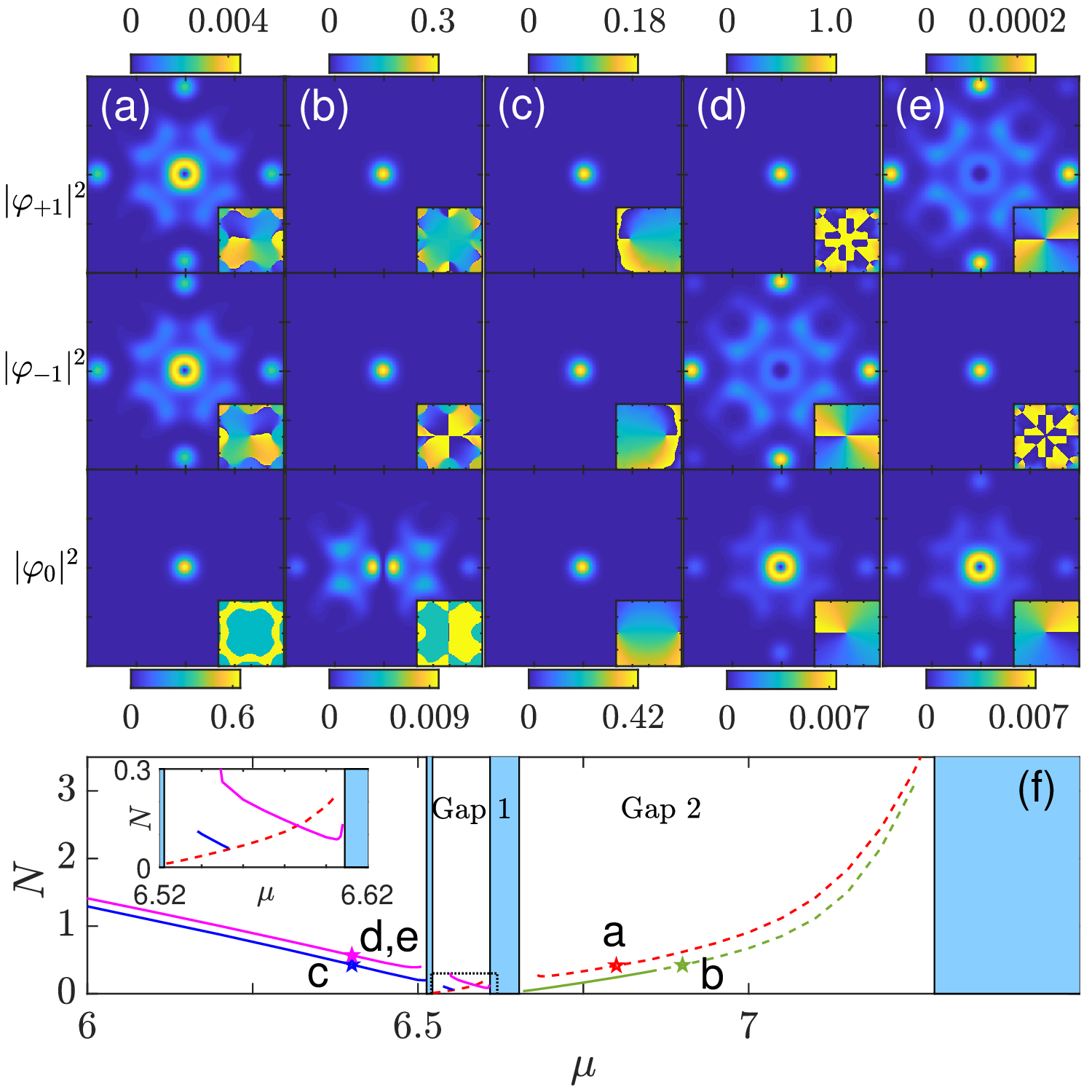}
	\caption{BT (Bloch-type) GS families bifurcating from the lowest two energy bands when $\theta=\arctan(3/4)$, $a=0.5\pi$, $V_0=4$, $\gamma=0.5$, $c_0=1$, and $c_2=-1.5$. (a)–(e) The density profiles of three components $\varphi_{+1}$ (the first row), $\varphi_{-1}$ (the second row), and $\varphi_{0}$ (the third row) for five GS families. The corresponding phases are shown in the subfigures. The first and second rows in (a)–(c) share the same color bars. $|\varphi_{-1}|$ in (d) and (e) are equal to $|\varphi_{+1}|$ in (e) and (d), respectively. Here, the drawing grid is $(x,y)\in [-4,4]$. (f) The norm $N$ vs the chemical potential $\mu$ for the five GS families. The norm curves for (d) and (e) are identical. The solid (dashed) parts of these curves indicate that the corresponding GSs are stable (unstable). The blue area represents the energy band, and the inset shows the magnified view of the region enclosed by the black dashed lines.}
	\label{4fig:4}
\end{figure}
As the ML period and lattice depth decrease, the low-energy linear bands gradually widen, and the energy gaps between them get narrow, until they collapse, leaving solely the semi-infinite gap. For example, in the case of $a=0.5\pi$ and $V_0=4$, the lowest two energy bands both exhibit finite bandwidth (this, being non-flat). Based on the norm curves of several WT GS families for different interaction strengths, we can identify GS families bifurcating from the lowest two energy bands in the case of a small ML period ($a=0.5\pi$), whose phases are similar to those of WT GS families ($a=2\pi$). Five GS families for $c_0=1$ and $c_2=-1.5$ are shown in Figs.~\ref{4fig:4}(a)-\ref{4fig:4}(e). These GSs span multiple unit cells, with envelopes modulated by the ML. The corresponding norm curves exhibit discontinuity while they hit energy bands, accompanied by a significant broadening of the envelope when $\mu$ approaches the band edges, as shown in Fig.~\ref{4fig:4}(f). This broadening effect intensifies with increasing bandwidth. The existence of a finite threshold, $N_{\mathrm{th}}=\min(N)\gg 0$, implies that these GSs cannot maintain localization and approach the Bloch states when the norm is very small. Therefore, these GSs do not admit the approximation by Wannier functions, while they bifurcate from the energy bands with finite bandwidth \cite{ShenS2024Two} and thus they are termed BT GSs. Nevertheless, these GS families still satisfy the relation similar to that given by Eq.~\eqref{4eq11}. The relation helps to find other BT GS families.

\subsection{Stability analysis of the gap solitons}
\begin{figure}[t]
	\centering
	\includegraphics[width=1\linewidth]{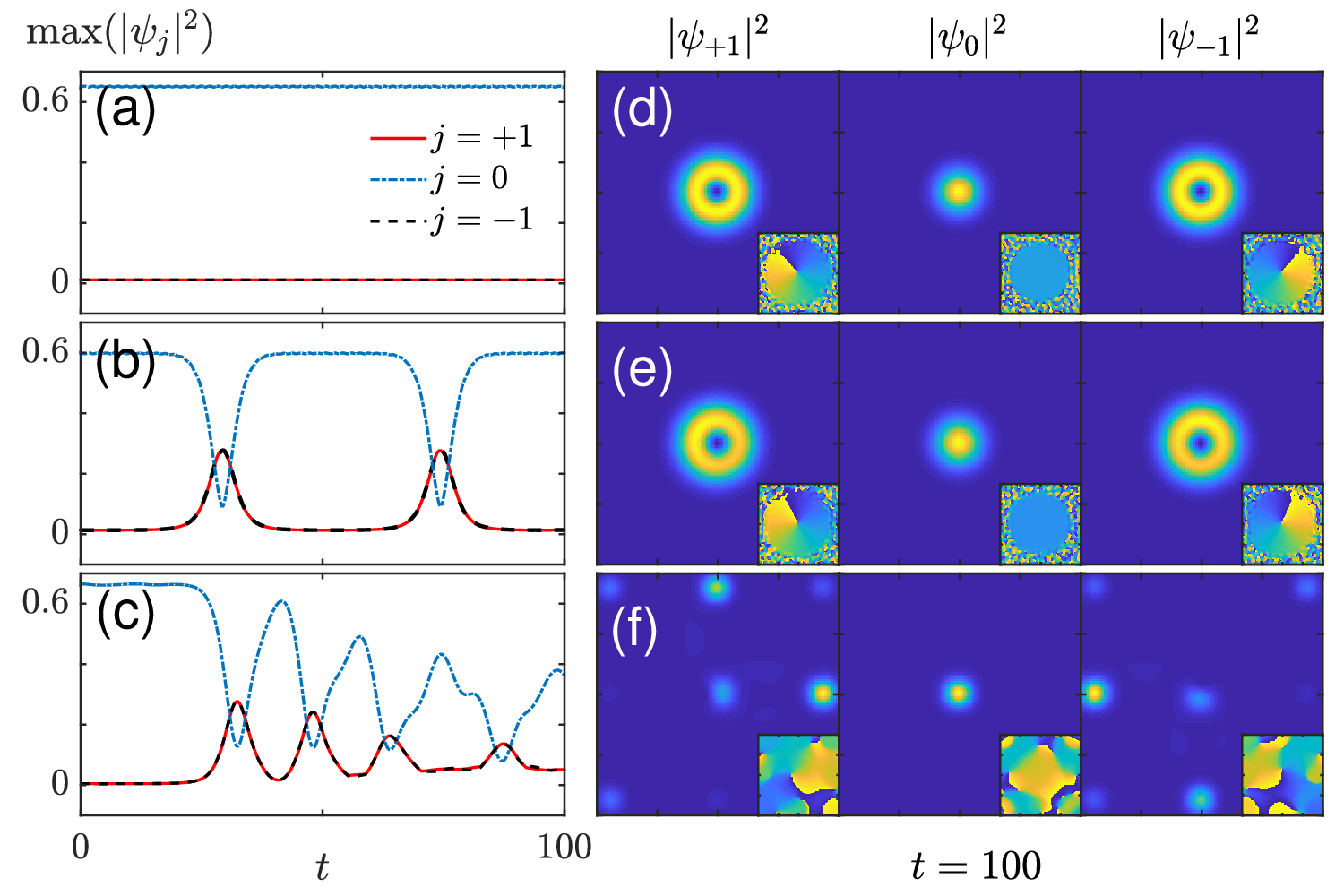}
	\caption{Panels (a)–(c) show the evolution of the maximum densities for the three components as revealed by the numerical simulations of the GSs in Figs. \ref{4fig:2}(a)-\ref{4fig:4}(a), respectively. The corresponding spatial density profiles of all three components at $t = 100$ are displayed in panels (d)–(f).}
	\label{4fig:5}
\end{figure}

Stability is the crucial characteristic of 2D solitons. We have systematically investigated the stability of the GSs families in the framework of the linearized Bogoliubov–de Gennes equations for small perturbations \cite{YangJ2010Nonlinear}. To this end, small perturbations $a_j(x,y)$ and $b_j(x,y)$ were added to the GS solutions as
\begin{equation}\label{4eq12}
\psi_j=\left(\varphi_j+a_je^{\lambda t}+b_j^*e^{\lambda^*t}\right)e^{-i\mu t},
\end{equation}
where $\lambda$ is the growth rate of the perturbations. Substituting the expression \eqref{4eq12} into Eq.~\eqref{4eq1} and performing the linearizing, we arrive at the eigenvalue equation $i\boldsymbol{L}\boldsymbol{\xi}=\lambda\boldsymbol{\xi}$, where $\boldsymbol{\xi}=[a_{+1},b_{+1},a_0,b_0,a_{-1},b_{-1}]^T$ and the matrix $\boldsymbol{L}$ is
\begin{widetext}
\begin{equation}
\begin{aligned}
\boldsymbol L&=\left[\begin{array}{cccccc}\label{4eq13}
L_{1} & -L_{11} & \frac{\gamma}{\sqrt{2}}\partial_{-}-L_{4} & -L_{12} & -L_{13} & -L_{6} \\
L_{11}^* & -L_{1} & L_{12}^* & \frac{\gamma}{\sqrt{2}}\partial_{+}+L_{4}^* & L_{6}^* & L_{13}^* \\
\frac{\gamma}{\sqrt{2}}\partial_{+}-L_{4}^* & -L_{12} & L_{2} & -L_{7} & \frac{\gamma}{\sqrt{2}}\partial_{-}-L_{5}^* & -L_{23} \\
L_{12}^* & \frac{\gamma}{\sqrt{2}}\partial_{-}+L_{4} & L_{7}^* & -L_{2} & L_{23}^* & \frac{\gamma}{\sqrt{2}}\partial_{+}+L_{5} \\
-L_{13}^* & -L_{6} & \frac{\gamma}{\sqrt{2}}\partial_{+}-L_{5} & -L_{23} & L_{3} & -L_{33} \\
L_{6}^* & L_{13} & L_{23}^* & \frac{\gamma}{\sqrt{2}}\partial_{-}+L_{5}^* & L_{33}^* & -L_{3}
\end{array}\right],
\end{aligned}
\end{equation}
\end{widetext}
with elements
\begin{equation}
\begin{aligned}
L_{1}&=L_0-C_+\left(2\rho_{+1}+\rho_{0}\right)-C_-\rho_{-1},\\
L_{2}&=L_0-C_+\left(\rho_{+1}+\rho_{-1}\right)-2c_0\rho_{0},\\
L_{3}&=L_0-C_+\left(2\rho_{-1}+\rho_{0}\right)-C_-\rho_{+1},\\
L_{11}&=C_+\varphi_{+1}^2,~~L_{4}=C_+\varphi_0^*\varphi_{+1}+2c_2\varphi_{-1}^*\varphi_{0},\\
L_{33}&=C_+\varphi_{-1}^2,~~L_{5}=C_+\varphi_0^*\varphi_{-1}+2c_2\varphi_{+1}^*\varphi_{0},\\
L_{12}&=C_+\varphi_0\varphi_{+1},~~L_{6}=C_-\varphi_{-1}\varphi_{+1}+c_2\varphi_0^2,\\
L_{23}&=C_+\varphi_{-1}\varphi_0,~~L_{7}=2c_2\varphi_{-1}\varphi_{+1}+c_0\varphi_0^2,\\
L_{13}&=C_-\varphi_{-1}^*\varphi_{+1},\\
\end{aligned}
\end{equation}
where $L_0=\nabla^2/2+\mu-V$, $C_+=c_0+c_2$, and $C_-=c_0-c_2$. By means of the numerical solution, we have obtained eigenvalues $\lambda$ corresponding to each stationary GS solution. Taking into account possible numerical errors, we define the GS to be stable provided that $\max\left(\mathrm{Re}(\lambda)\right)<10^{-3}$. The so established stability was verified through numerical simulations of the perturbed GSs. The stability of the above-found GS families is designated, respectively, by continuous and dashed segments of the corresponding norm curves in Figs. \ref{4fig:2}(f)-\ref{4fig:4}(f). It is seen that the WT GSs are stable only when $N$ is sufficiently small. Note that among the GS families that branch off to the left (right) from the bifurcation point, those with non zero (zero) vorticity have larger stability domains. The BT GSs exhibit stability properties similar to those of their WT GSs counterparts. Note that the BT GSs are unstable as they approach edges of the energy bands, which is essential as the energy bands widen, in particular, with the reduction of the lattice depth $V_0$. We performed extensive numerical simulations by using numerical solutions with random perturbations added to the input. The results, which are consistent with the linear stability analysis, are exemplified by three cases corresponding to the GSs shown in Figs. \ref{4fig:2}(a)-\ref{4fig:4}(a), as shown in Fig. \ref{4fig:5}.
Stable GSs exhibit long-term robustness, whereas unstable ones rapidly decay.

\subsection{The transition between Wannier-type and Bloch-type gap solitons induced by spin-orbit coupling}
The WT (BT) GS families bifurcate from the nearly flat (strongly curved) bands. The key distinction between the two GS species is manifested by the fact if the wave functions remain localized at $N\rightarrow 0$, i.e., whether GSns with small norms exist. Notably, Fig.~\ref{4fig:1}(f) reveals that the SOC strength $\gamma$ affects the flatness of the energy bands, thus determining the GS type, WT or BT. We quantify the flatness of the energy bands (with index $n$) by a parameter
\begin{equation}\label{4eq15}
\Delta_n=\max(\mu_n)-\min(\mu_n),
\end{equation}
with smaller $\Delta_n$ indicating a flatter energy band.

\begin{figure}[t]
	\centering
	\includegraphics[width=1\linewidth]{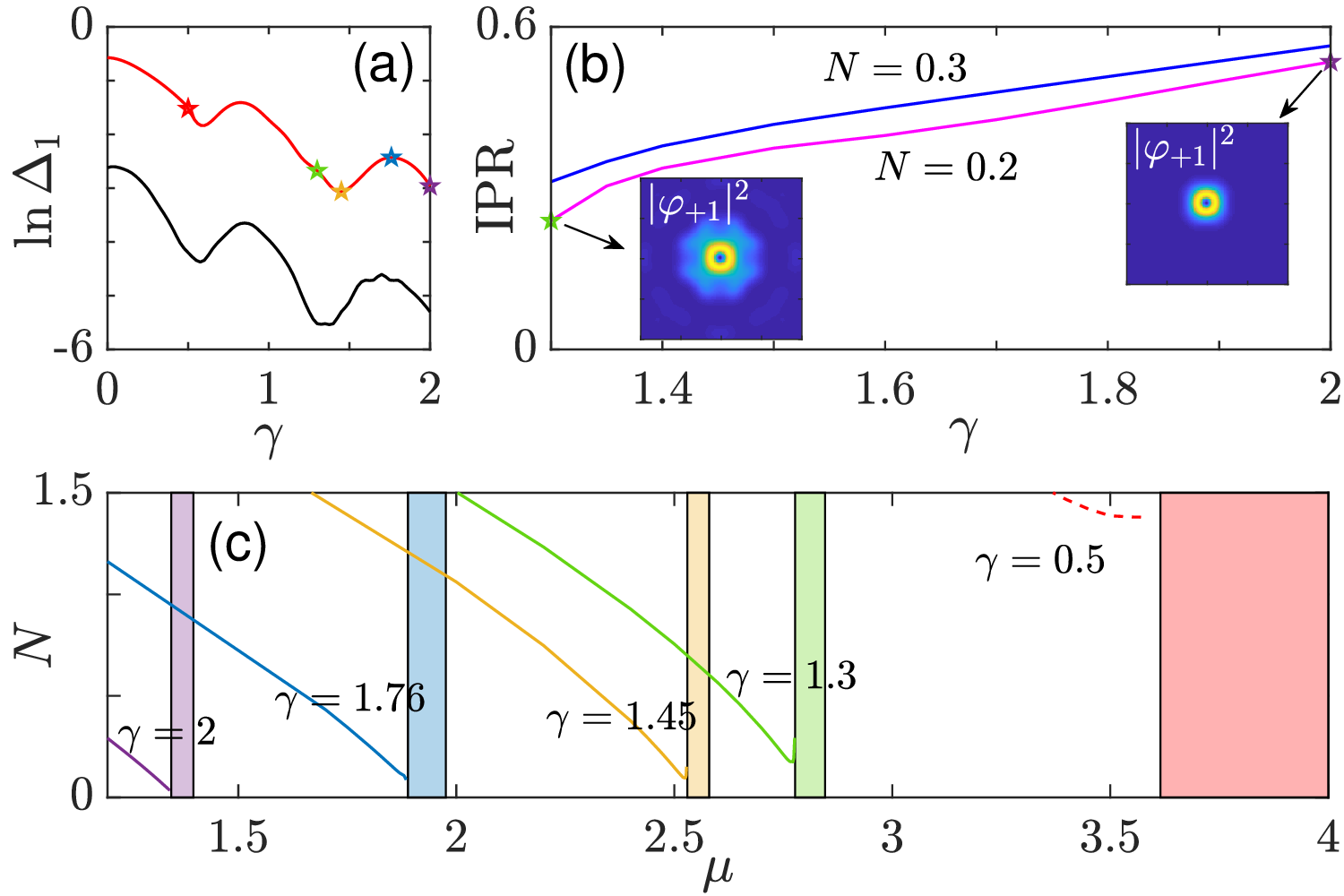}
	\caption{(a) The effect of the SOC strength $\gamma$ on the flatness $\Delta_1$ of the lowest-energy band for $V_0=2$ (the red line) and $V_0=4$ (the black line). (b) The inverse participation ratio (IPR) of the GS solutions bifurcating from the lowest-energy band vs $\gamma$ for different values of $N$ when $V_0=2$. The corresponding subfigures show the density profiles of the component $\varphi_{+1}$ at $\gamma=1.3$ and $\gamma=2$. Panel (c) displays the corresponding norm curves and the lowest-energy bands for the respective values of $\gamma$ marked in (a). The solid (dashed) parts of these curves indicate that the corresponding GSs are stable (unstable). Here, $\theta=\arctan(3/4)$, $a=0.5\pi$, $c_0=-1$, and $c_2=1.5$.}
	\label{4fig:6}
\end{figure}

Here we focus on the lowest-energy band ($n=1$). The effect of the SOC strength $\gamma$ on the parameter $\Delta_1$, defined as per Eq.~\eqref{4eq15}, is shown in Fig.~\ref{4fig:6}(a) for $V_0=2$ (the red line) and $V_0=4$ (the black line), with other parameters set as $\theta=\arctan(3/4)$ and $a=0.5\pi$.
We find that increasing $\gamma$ leads to a periodic decrease and increase in the flatness of the lowest-energy band, but overall, $\Delta_1$ exhibits a decreasing trend. To systematically investigate the effect of the SOC strength on GSs, we obtain the GS solutions bifurcating from the lowest-energy band at a fixed norm for different values of $\gamma$. The results show that as $\gamma$ increases, the soliton profile narrows and its amplitude grows, indicating enhanced localization. This trend is clearly reflected both in the inverse participation ratio (IPR),
\begin{equation}\label{4eq16}
\mathrm{IPR}=\frac{1}{N^2}\int_{-\infty}^{+\infty}\left(\sum_{j=-1}^{j=+1}|\varphi_j|^2\right)^2\mathrm{d}x\mathrm{d}y,
\end{equation}
which quantifies the degree of localization, and in a direct comparison of the density distributions at $\gamma=1.3$ and $\gamma=2$, as shown in Fig.~\ref{4fig:6}(b). The focusing effect induced by $\gamma$ is also the reason why SOC can stabilize two-dimensional solitons in the absence of an external potential. Furthermore, the norm curves of GS families are compared for $\gamma=0.5$, $\gamma=1.3$, $\gamma=1.45$, $\gamma=1.76$, and $\gamma=2$ in Fig.~\ref{4fig:6}(c). These results show a distinct transition in the character of the norm curves: the monotonicity evolves from non monotonic to monotonic as $\gamma$ increases, accompanied by a reduction in the norm threshold. Both the change in monotonicity and the reduction in norm threshold consistently indicate a gradual transition of the soliton type toward the Wannier type. Therefore, we conclude that for appropriately selected lattice parameters, an increase in $\gamma$ can enhance the band flatness to the level necessary for the emergence of WT GSs, driving the transition of BT GSs to WT GSs.

\subsection{Soliton families in quasiperiodic moir\'e lattices}
When the twist angle $\theta$ is not a Pythagorean angle, the ML $V(x,y)$ becomes quasiperiodic (with an infinite period and a zero-area Brillouin zone). It can be approximated by a periodic ML corresponding to a nearby Pythagorean angle $\theta^{\prime}$, where the approximation accuracy can theoretically be improved indefinitely. Below we briefly describe the periodic approximation method for a non-Pythagorean twist angle $\theta=\pi/4$.

\begin{figure}[t]
	\centering
	\includegraphics[width=1\linewidth]{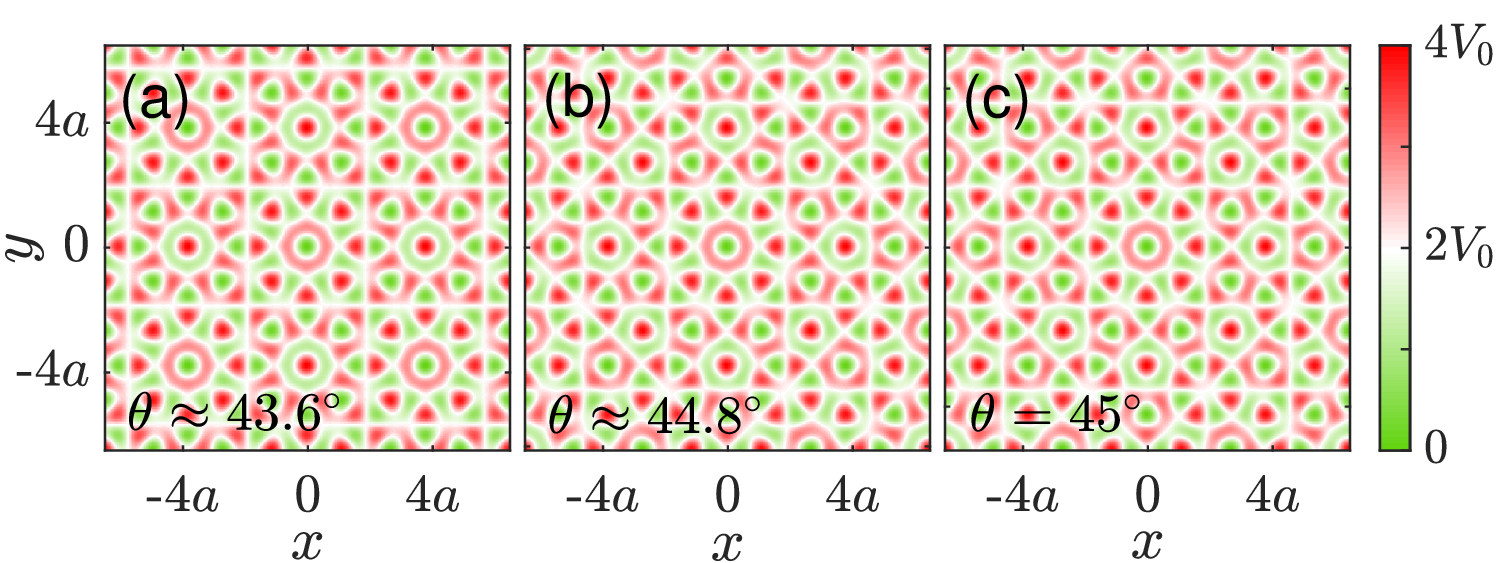}
	\caption{The ML profiles for two Pythagorean angles, $\theta=\arctan(20/21)\approx 43.6^{\circ}$ (a), $\theta=\arctan(119/120)\approx 44.8^{\circ}$ (b), and the non-Pythagorean one, $\theta=\pi/4$ (c).}
	\label{4fig:7}
\end{figure}

For a Pythagorean angle $\theta^{\prime}$, the corresponding Pythagorean triple ($\bar{a},\bar{b},\bar{c}$) can be represented by two coprime natural numbers ($\bar{m},\bar{n}$), namely, $\bar{a}=\bar{m}^2-\bar{n}^2$, $\bar{b}=2\bar{m}\bar{n}$, and $\bar{c}=\bar{m}^2+\bar{n}^2$. By solving $\sin\theta\approx\sin\theta^{\prime}=(\bar{m}^2-\bar{n}^2)/(\bar{m}^2+\bar{n}^2)$, all Pythagorean angles that approximate a given non-Pythagorean angle $\theta$ can be derived. For $\theta=\pi/4$, the two coprime natural numbers satisfy a ratio condition $\bar{m}/\bar{n}=\sqrt{3+2\sqrt{2}}\approx2.41$. The approximate solutions ($\bar{m},\bar{n}$) include $(5,2)$, $(12,5)$, etc., and the corresponding Pythagorean triples ($\bar{a},\bar{b},\bar{c}$) are $(21,20,29)$ and ($119,120,169$), respectively. In Figs. \ref{4fig:7}(a)-\ref{4fig:7}(c),
the direct comparison of the ML profiles for the non-Pythagorean angle ($\theta=\pi/4$) and two Pythagorean angles [$\theta^{\prime}=\arctan(20/21)$ and $\theta^{\prime}=\arctan(119/120)$] clearly demonstrates the accuracy of the periodic approximation method.

Quasiperiodic potentials enable the formation of localized states in purely linear systems through the Anderson-localization mechanism \cite{Hiramoto1989New}. A natural question is what happens to 2D solitons under the action of quasiperiodic potentials. Here, we address this issue in the case of $\theta=\pi/4$, not considering the edge states induced by finite boundaries \cite{KuzmenkoI2024Fermionic}. For the large lattice period $a=2\pi$, the quasiperiodic ML still supports all bright GS families shown in Figs. \ref{4fig:2} and \ref{4fig:3}. Profiles of these solitons remain nearly identical to those in the periodic ML with $\theta=\arctan(3/4)$, and the corresponding norm curves nearly coincide. The point is that these solitons are essentially localized in the central parabolic-like potential when $a=2\pi$, rendering the effective models, and solitons produced by the models, nearly indistinguishable for Pythagorean and non-Pythagorean twist angles.

\begin{figure}[t]
	\centering
	\includegraphics[width=1\linewidth]{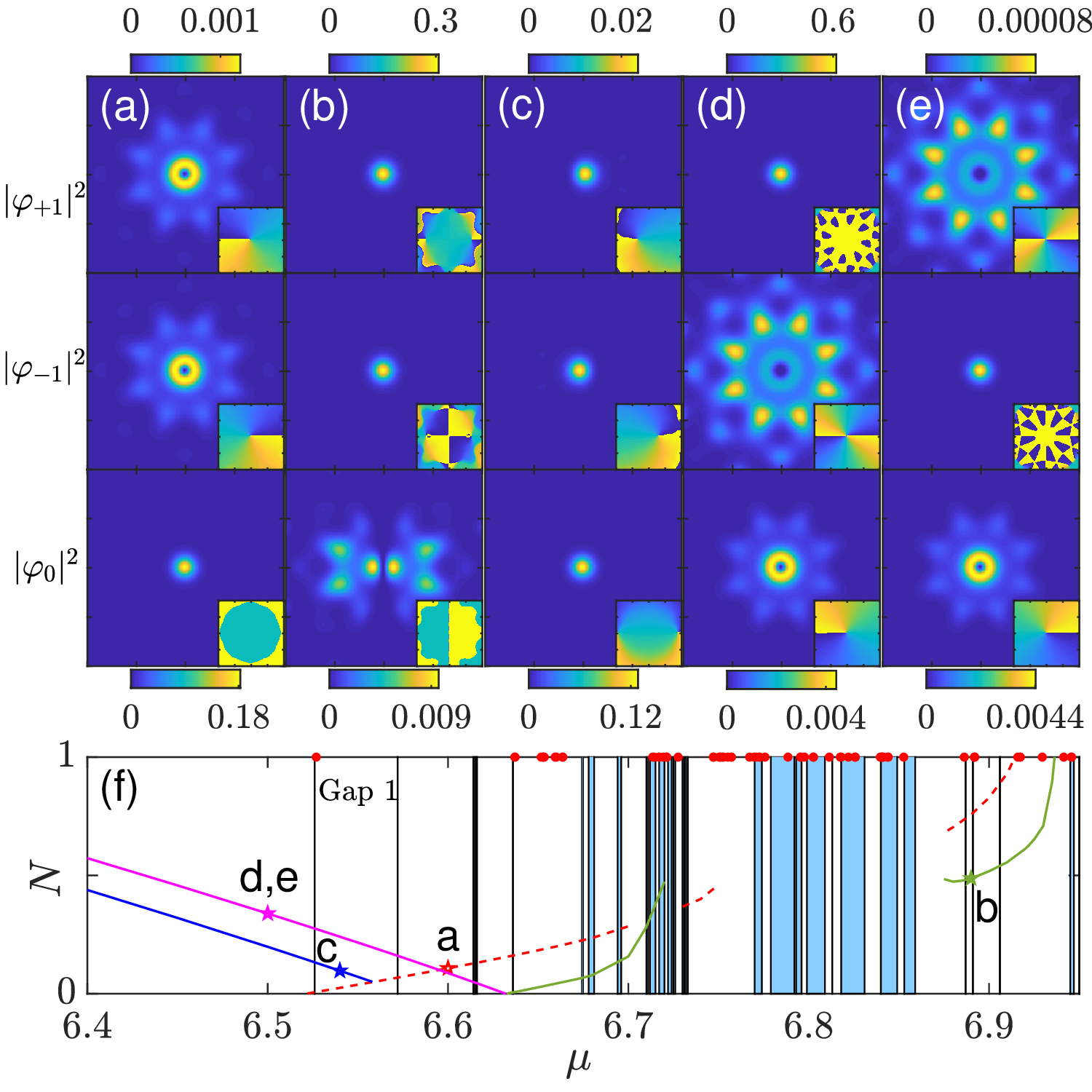}
	\caption{GS families for the non-Pythagorean angle $\theta=\pi/4$ when $a=0.5\pi$, $V_0=4$, $\gamma=0.5$, $c_0=1$, and $c_2=-1.5$. (a)–(e) The density profiles of three components $\varphi_{+1}$ (the first row), $\varphi_{-1}$ (the second row), and $\varphi_{0}$ (the third row) for five GS families. The corresponding phases are shown in the subfigures. The first and second rows in (a)–(c) share the same color bars. $|\varphi_{-1}|$ in (d) and (e) are equal to $|\varphi_{+1}|$ in (e) and (d), respectively. The drawing grid is $(x,y)\in [-4,4]$. (f) Norm $N$ vs the chemical potential $\mu$ for the five GS families. The norm curves for (d) and (e) are identical. The solid (dashed) parts of these curves indicate that the corresponding GSs are stable (unstable). The red dots represent the linear spectrum for the ML with $\theta=\pi/4$. The blue areas represent the energy bands for the ML with $\theta^{\prime}=\arctan(119/120)$.}
	\label{4fig:8}
\end{figure}

Reducing the lattice period breaks the similarity of the effective models for different twist angles. For $\theta=\pi/4$, $a=0.5\pi$, and $V_0=4$, we further identify five bright soliton families phase-matched to those in Fig. \ref{4fig:4}, with parameters $\gamma=0.5$, $c_0=1$, and $c_2=-1.5$, as shown in Figs. \ref{4fig:8}(a)-\ref{4fig:8}(e).
Notably, the GSs supported by the ML with $\theta=\pi/4$ exhibit significantly different density distributions from those at $\theta=\arctan(3/4)$, but they still satisfy the approximate linear relation similar to that described in Eq. \eqref{4eq11}. The numerically generates norm curves of five GS families are plotted in Fig. \ref{4fig:8}(f), including their stability. The norm curves for soliton families in Figs. \ref{4fig:8}(c)-\ref{4fig:8}(e) maintain continuity, while the curves for soliton families in Figs. \ref{4fig:8}(a) and \ref{4fig:8}(b) exhibit two discontinuities (where the soliton does not exist), indicating the presence of continuous energy spectra (without gaps) in the two regions. To verify this, we directly compute the linear spectrum (red dots) of the quasiperiodic system at $\theta=\pi/4$ alongside the band-gap structure (blue areas) of the periodic system at $\theta^{\prime}=\arctan(119/120)$, as shown in Fig. \ref{4fig:8}(f). In the linear spectrum, the two isolated red dots on the left precisely locate bifurcation points of the soliton families, as do the two flat bands in the band-gap structure. Moreover, regions with dense sets of eigenvalues in the linear spectrum and broad bands in the band-gap structure clearly display the discontinuous parts of the norm curves, where solitons are absent. These inferences further validate the accuracy of the norm curves obtained by the numerical method. The comparison of Figs. \ref{4fig:4}(f) and \ref{4fig:8}(f) reveals that the stability is primarily governed by the type of the soliton, independent of twist angles. On the other hand, we find SOC can also enhance the localization of the solitons in quasiperiodic MLs.

\section{Conclusions}\label{4-4}
We have systematically investigated GSs (gap solitons) in the dynamical models of the spin-1 BECs under the action of the Rashba SOC (spin-orbit coupling) in a periodic or quasiperiodic ML (moir\'e lattice). We focused on the GSs bifurcating from the five lowest-energy bands. GSs in periodic systems are classified into two species, WT (Wannier-type) and BT (Bloch-type), according to their ability (WT) or inability (BT) to maintain the localization at vanishingly small norms. When the ML depth and period are sufficiently large, the lowest five energy bands become flat and five WT GS families bifurcate from them, each approximated by a single Wannier function. More complex WT GSs are also constructed via the coherent superposition of the fundamental WT GSs with equal chemical potentials. The WT GSs are stable only at small norms. Conversely, when the ML depth and period are small, the energy bands exhibit finite bandwidths. Guided by the bifurcation characteristics of WT GSs, we have identified BT GSs whose phase patterns are similar to those of WT GSs. The broad energy bands cause the width of BT GSs to significantly expand near band edges, thereby inducing instability. Additionally, the Rashba SOC effect has been demonstrated to enhance the soliton localization while reducing the norm threshold above which the GSs exist. These findings suggest that, even with small ML depth and period, strongly localized WT GSs can still be generated by increasing the SOC strength. We have also established that quasiperiodic MLs affect the spatial profiles and existence domains of the GSs without altering their core stability mechanisms. The edge states induced by finite boundaries in quasiperiodic MLs also constitute a promising topic for future investigation. Overall, our results offer a theoretical basis for relevant experimental investigations and propose a systematic approach to studies of GSs in complex periodic and quasiperiodic systems.

\begin{acknowledgments}
	We acknowledge support of the National Natural Science Foundation of China (Grants No.~11835011, No.~12375006, and No.~12074343) and the Natural Science Foundation of Zhejiang Province of China (Grant No.~LZ22A050002).
\end{acknowledgments}

\end{document}